\documentclass[letterpaper]{article} 
\usepackage{aaai25}  
\usepackage{times}  
\usepackage{helvet}  
\usepackage{courier}  
\usepackage[hyphens]{url}  
\usepackage{graphicx} 
\usepackage{natbib}  
\usepackage{caption} 
\usepackage{algorithm}
\usepackage{algorithmic}
\usepackage{booktabs}
\usepackage{multirow}
\usepackage{xcolor}
\usepackage{amsmath}
\usepackage{utfsym}
\usepackage{epsfig}
\usepackage{graphicx}
\usepackage{amssymb}
\usepackage{dblfloatfix}
\usepackage{transparent}

\usepackage{algorithm}
\usepackage{listings}

\usepackage{etoolbox}
\makeatletter
\AfterEndEnvironment{algorithm*}{\let\@algcomment\relax}
\AtEndEnvironment{algorithm*}{\kern2pt\hrule\relax\vskip3pt\@algcomment}
\AfterEndEnvironment{algorithm}{\let\@algcomment\relax}
\AtEndEnvironment{algorithm}{\kern2pt\hrule\relax\vskip3pt\@algcomment}
\let\@algcomment\relax
\newcommand\algcomment[1]{\def\@algcomment{\footnotesize#1}}
\renewcommand\fs@ruled{\def\@fs@cfont{\bfseries}\let\@fs@capt\floatc@ruled
  \def\@fs@pre{\hrule height.8pt depth0pt \kern2pt}%
  \def\@fs@post{}%
  \def\@fs@mid{\kern2pt\hrule\kern2pt}%
  \let\@fs@iftopcapt\iftrue}
\makeatother

\newcommand{\Rmnum}[1]{\uppercase\expandafter{\romannumeral #1}}  

\usepackage{newfloat}
\usepackage{listings}
\DeclareCaptionStyle{ruled}{labelfont=normalfont,labelsep=colon,strut=off} 
\floatstyle{ruled}
\newfloat{listing}{tb}{lst}{}
\floatname{listing}{Listing}
\nocopyright
\title{HLLM: Enhancing Sequential Recommendations via Hierarchical Large Language Models for Item and User Modeling}
\author{
    Junyi Chen\thanks{Equal contribution. $^\text{\textdagger}$Corresponding author.},
    Lu Chi\footnotemark[1],
    Bingyue Peng,
    Zehuan Yuan$^\text{\textdagger}$
}
\affiliations{
    ByteDance \\
    \{chenjunyi.s, chilu, bingyue.peng, yuanzehuan\}@bytedance.com
}

\usepackage{bibentry}

\begin{document}

\maketitle

\begin{abstract}
Large Language Models (LLMs) have achieved remarkable success in various fields, prompting several studies to explore their potential in recommendation systems. However, these attempts have so far resulted in only modest improvements over traditional recommendation models. Moreover, three critical questions remain under-explored: firstly, the real value of LLMs' pre-trained weights, often considered to encapsulate world knowledge; secondly, the necessity of fine-tuning for recommendation tasks; lastly, whether LLMs can exhibit the same scalability benefits in recommendation systems as they do in other domains.
In this paper, we propose a novel \textbf{H}ierarchical \textbf{L}arge \textbf{L}anguage \textbf{M}odel (HLLM) architecture designed to enhance sequential recommendation systems. Our approach employs a two-tier model: the first Item LLM extracts rich content features from the detailed text description of the item, while the second User LLM utilizes these features to predict users' future interests based on their interaction history.
Extensive experiments demonstrate that our method effectively leverages the pre-trained capabilities of open-source LLMs, and further fine-tuning leads to significant performance boosts. Additionally, HLLM achieves excellent scalability, with the largest configuration utilizing 7B parameters for both item feature extraction and user interest modeling. 
Moreover, HLLM offers excellent training and serving efficiency, making it practical in real-world applications.
Evaluations on two large-scale datasets, PixelRec and Amazon Reviews, show that HLLM achieves state-of-the-art results, outperforming traditional ID-based models by a wide margin. 
In online A/B testing, HLLM showcases notable gains, validating its practical impact in real-world recommendation scenarios. Codes are available at \url{https://github.com/bytedance/HLLM}.
\end{abstract}

%

\section{Introduction}

The recommendation algorithm is a classic yet complex problem that requires understanding user interests to predict future behaviors across various items. The key to effective recommendation lies in accurately modeling both item and user features. Currently, mainstream approaches are predominantly ID-based, converting items and users into IDs and creating corresponding embedding tables for encoding~\cite{goldberg1992using,koren2009matrix,sarwar2001item}. 
To capture diverse and temporally varying user interests, several sequential modeling methods have been developed, demonstrating notable success in sequential recommendations~\cite{hidasi2015session,zhou2018deep,sasrec,sun2019bert4rec}. However, these methods are typically dominated by embedding parameters and have relatively small model sizes, leading to two major drawbacks: a heavy reliance on ID features which results in poor performance in cold-start scenarios, and relatively shallow neural networks find it difficult to model complex and diverse user interests.

With the advent of ChatGPT~\cite{chatgpt}, large language models (LLMs) have achieved significant breakthroughs across various domains, showcasing impressive world knowledge and reasoning capabilities~\cite{touvron2023llama,achiam2023gpt,team2023gemini}. This success has spurred interest among researchers in exploring the integration of LLMs into recommendation systems~\cite{wu2023survey,li2023large}. These explorations can be broadly categorized into three approaches: (1). Utilizing LLMs to provide refined or supplementary information for recommendation systems~\cite{zhang2024spar,ren2024representation,xi2023towards}, such as summary of user behavior and item information expansion. (2). Transforming the recommendation system into a dialogue-driven format compatible with LLMs~\cite{bao2023tallrec,friedman2023leveraging,zhang2023recommendation,yang2023palr,zhai2023knowledge}. (3). Modifying LLMs to handle recommendation tasks beyond just text input and output. This includes approaches that input ID features into LLMs~\cite{ning2024user,meta,liao2024llara} and those that replace existing models with LLMs, optimizing directly for objectives like Click-Through Rate (CTR)~\cite{cui2022m6,kang2023llms}.

Despite these advancements, integrating LLMs with recommendation systems presents notable challenges in complexity and effectiveness. One issue is that inputting user behavior history as text to LLMs results in very long input sequences. Consequently, LLMs need longer sequences to represent the same time span of user behavior than ID-based methods, while the complexity of the self-attention module in LLMs scales quadratically with the sequence length. 
Additionally, recommending a single item requires generating several text tokens, leading to multiple forwards and resulting in lower efficiency.
In terms of effectiveness, the performance improvements of existing LLM-based methods over traditional methods are not significant, raising questions about whether the potential of LLMs has been fully realized.

Moreover, some critical issues remain underexplored. Firstly, the actual value of pre-trained LLM weights, often regarded as encapsulating world knowledge, needs further investigation. While LLMs offer impressive zero-shot and few-shot capabilities, their value when training on large-scale recommendation data is unclear. Secondly, the necessity of fine-tuning for recommendation tasks is in question. LLMs pre-trained on massive corpora exhibit strong world knowledge, but whether further fine-tuning on recommendation tasks enhances or diminishes performance remains to be seen. Lastly, the scalability of LLMs, a hallmark characteristic with proven scaling laws in other domains, requires validation in the context of recommendation systems. While some studies have successfully validated the scaling laws in the recommendation domain~\cite{shin2023scaling, meta}, these models have considerably fewer parameters compared to LLMs. Whether models exceeding 1 billion parameters exhibit good scalability in the recommendation domain remains an open question.

To address these challenges, this paper proposes the \textbf{H}ierarchical \textbf{L}arge \textbf{L}anguage \textbf{M}odel (HLLM) architecture. The approach begins by using an LLM to extract item features. To empower the LLM to effectively extract these features, a special token is appended to the end of the detailed textual description of each item. This augmented description is then input into the LLM (referred to as the Item LLM), and the output corresponding to the special token is used as the item feature. These item features are then input into a second LLM (referred to as the User LLM) to model user interest and predict future behaviors. By transforming extensive item descriptions into concise embeddings, the length of behavior sequences is reduced to that of ID-based models, significantly lowering computational complexity compared to other text-based LLM recommendation models. We also verified that HLLM has a significant training efficiency advantage compared to ID-based models, as it can surpass ID-based models with only a small amount of training data.

Extensive experiments are conducted to explore the value of pre-training. Although the HLLM does not employ text interaction in the conventional manner of standard LLMs, such as the Item LLM being designed as a feature extractor, and both input and output of the User LLM being item embeddings, the pre-trained weights have proven beneficial for both types of LLMs. This demonstrates that the world knowledge embedded in LLMs is indeed valuable for recommendations. Nevertheless, this does not obviate the need for fine-tuning towards recommendation objectives. Conversely, our experiments indicate that such fine-tuning is crucial for surpassing traditional methods. To verify scalability, experiments on large academic datasets confirm that LLMs exhibit excellent scalability with performance improving as model parameters increase. Within the limited resources, models up to 7 billion parameters show consistent performance gains with increasing size.

Ultimately, the proposed HLLM architecture outperforms existing methods across multiple academic datasets, achieving state-of-the-art results. More importantly, the effectiveness of HLLM is also validated through real-world online A/B testing, confirming its practical applicability.

Our main contributions can be summarized as follows:

1) A novel hierarchical LLM (HLLM) framework is introduced for sequential recommendations. This approach significantly outperforms classical ID-based models on large-scale academic datasets and has been validated to yield tangible benefits in real-world industrial settings. Additionally, this method demonstrates excellent training and serving efficiency.

2) HLLM effectively transfers the world knowledge encoded during the LLM pre-training stage into the recommendation model, encompassing both item feature extraction and user interest modeling. Nevertheless, task-specific fine-tuning with recommendation objectives is essential.

3) HLLM exhibits excellent scalability, with performance continuously improving as the data volume and model parameters increase. This scalability highlights the potential of the proposed approach when applied to even larger datasets and model sizes.

\section{Related Work}

\subsection{Traditional Recommender Systems}
Traditional Recommender Systems predominantly rely on ID-based embeddings, and how to design feature interactions is an important topic. DeepFM~\cite{guo2017deepfm} models low-order feature interactions with FM and models high-order feature interactions with DNN. DCN~\cite{wang2017deep,wang2021dcn} can model higher-order interactions by explicitly applying feature crossing at each layer. Besides, some researchers make efforts to model user interests from their historical behavior. For instance, DIN~\cite{zhou2018deep} and DIEN~\cite{zhou2019deep} introduce attention mechanisms to capture user's diverse interests from historical behaviors. Inspired by transformer, SASRec~\cite{sasrec} applies self-attention
mechanisms to sequential recommendation. CLUE~\cite{shin2023scaling} and HSTU~\cite{meta} demonstrate that models with parameter counts within hundreds of millions adhere to the scaling law. Some works have also introduced content features into recommendation models, showing certain advantages in generalization~\cite{baltescu2022itemsage,li2023text,pixelrec}.

\subsection{Recommendation with Language Models}
The success of LLMs has attracted many researchers to explore their applications in recommendation systems. These explorations can be categorized into three types. Firstly, LLMs are used for summarizing or supplementing information about users or items~\cite{zhang2024spar,ren2024representation,xi2023towards}. For example, RLMRec~\cite{ren2024representation} develops a user/item profiling paradigm empowered by LLMs, and aligns the semantic space of LLMs with the representation space of collaborative relational signals through a cross-view alignment framework. LLMs are also employed to generate augmented training signals for coldstart items~\cite{wang2024large}. 
Secondly, some works adapt the recommendation domain data into conversational formats~\cite{bao2023tallrec,friedman2023leveraging,zhang2023recommendation,yang2023palr,zhai2023knowledge}.
\begin{figure}
    \centering
    \includegraphics[width=.9\linewidth]{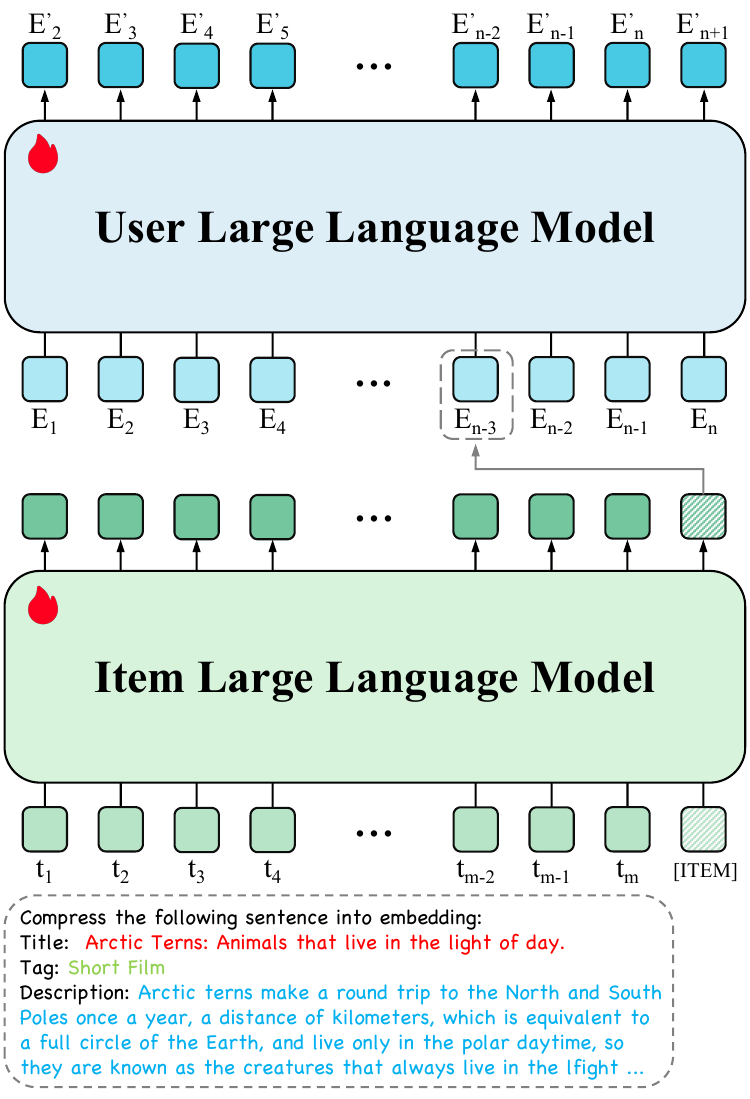}
    \caption{Architecture of Hierarchical Large Language Model.
    HLLM consists of two LLMs with non-shared parameters: Item LLM and User LLM. The Item LLM takes the text description of an item as input, appended with a special token \texttt{[ITEM]}, and outputs the item embedding. The User LLM inputs the item embeddings of the user's historical interactions and predicts next item. All LLM parameters are trainable and optimized via next item prediction.
    }
    \label{HLLM}
    \vspace{-0.3cm}
\end{figure}
Some approaches treat the recommendation task as a special form of instruction-following, inputting user historical behaviors in text form to the LLM to predict subsequent actions~\cite{li2024calrec}. Lastly, there are also some works that have adapted LLMs for recommendation tasks, allowing their inputs or outputs to go beyond just textual forms. LLaRA~\cite{liao2024llara} proposed a novel hybrid prompting method that integrates ID-based item embeddings with textual item features. LEARN~\cite{jia2024knowledge} utilizes pre-trained LLMs to extract item features. LLMs are also adapted to multi-class classification or regression for rating prediction~\cite{kang2023llms}. However, these methods offer limited improvements compared to traditional recommendation models.

\section{Method}

In this section, we first introduce the problem formulation, and then propose \textbf{H}ierarchical \textbf{L}arge \textbf{L}anguage \textbf{M}odel (HLLM) with a detailed explanation of how to adapt pre-trained large language models to recommendation systems, including item feature extraction and user interest modeling. Finally we discuss how to align HLLM with the objectives of recommendation systems, thereby significantly enhancing its performance on recommendation tasks.

\subsection{Problem Formulation}
We study the task of sequential recommendations, formulated as: Given a user $u \in \mathcal{U}$, a sequence of user $u$'s historical interactions $U = \{I_1, I_2, \dots, I_n\}$ in chronological order, predict the next item $I_{n+1}$, where $n$ is the length of $U$ and $I \in \mathcal{I}$. Each item $I$ has its corresponding ID and text information (e.g. title, tag, etc.), but the method proposed in this paper uses only the text information.

\subsection{Hierarchical Large Language Model Architecture}

Currently, a considerable number of LLM-based recommendation models flatten users' historical behaviors into plain text inputs for the LLM~\cite{kang2023llms,yang2023palr,li2024calrec}. This results in very long input sequences, and due to the self-attention module in LLMs, the complexity grows quadratically with the length of the input sequence. To reduce the burden of user sequence modeling, we adopt a hierarchical modeling approach called the \textbf{H}ierarchical \textbf{L}arge \textbf{L}anguage \textbf{M}odel (HLLM) that decouples item modeling from user modeling, as shown in Figure~\ref{HLLM}. Specifically, we first extract item features using the Item LLM, compressing the complex text descriptions into an embedding representation. Then, we model the user profile based on these item features with the User LLM. Additionally, to ensure better compatibility with pre-trained LLMs and to enhance scalability, we introduce minimal structural changes and design simple yet efficient training objectives. The following is a detailed introduction to item and user modeling.

\subsubsection{Item LLM}
is proposed to extract item features. It takes as input the text description of an item and outputs an embedding representation. LLMs have demonstrated excellent performance in text comprehension, but their use has mostly been limited to text generation scenarios, with few works using them as feature extractors. Inspired by previous works~\cite{devlin2018bert,neelakantan2022text}, a special token \texttt{[ITEM]} is added at the end of the item's text description to extract features.

Specifically, as shown in Figure~\ref{HLLM}, for Item $I$ we first flatten its corresponding textual attributes into the sentence $T$, and prepend it with a fixed prompt. 
After passing through the LLM tokenizer, we additionally append a special token \texttt{[ITEM]} at the end, thus the input token sequence for the Item LLM can be formulated as $\{t_1, t_2, \dots, t_m, \texttt{[ITEM]}\}$ where $m$ represents the length of text tokens. The hidden state from the last layer corresponding to the special token \texttt{[ITEM]} is considered as the item embedding.

\subsubsection{User LLM}
is designed to model user interests which is another key aspect of recommendation systems. 
The original user history sequence $U = \{I_1, I_2, \dots, I_n\}$ can be transformed into a historical feature sequence $\{E_1, E_2, \dots, E_n\}$ through the Item LLM, where $E_i$ represents the item embedding of $I_i$. 
The User LLM takes this historical feature sequence as input and predict next item embedding based on a sequence of previous interactions. As shown in Figure~\ref{HLLM}, the output of the User LLM corresponding to $E_i$ is $E_{i+1}^\prime$, which is expected to be the embedding of $I_{i+1}$.

Unlike traditional LLMs with text-in and text-out formats, here both the input and output of the User LLM are item embeddings. Therefore, we discard the word embeddings from the pre-trained LLM but retain all other pre-trained weights. Experiments show that these pre-trained weights are very helpful for reasoning user interests.

\subsection{Training for Recommendation Objectives}

Existing LLMs are all pre-trained using general natural language corpora. Although they possess a wealth of world knowledge and strong reasoning abilities, there remains a considerable gap between their capabilities and those required by recommendation systems. Following the best practices of other works~\cite{zhou2024lima,touvron2023llama}, we adopt supervised fine-tuning on top of the pre-trained LLM.

Recommendation systems can be divided into two categories, generative and discriminative recommendation. It is noteworthy that the proposed HLLM architecture is applicable to both types, requiring only appropriate adjustments to the training objectives. The following sections provide a detailed introduction to the training objectives for both categories.

\subsubsection{Generative Recommendation}

Recent work~\cite{meta} has provided a successful generative recommendation solution, including both retrieval and ranking. Our approach differs from it in two major ways: the model architecture is upgraded to large language models with pre-trained weights, and the input features are changed from IDs to text-input LLM features. The above differences have minimal impact on the training and serving strategies, therefore, we largely follow approaches proposed in~\cite{meta}.

For the training objective of generative recommendation, next item prediction is adopted, which aims to generate the embedding of the next item given the embeddings of the previous items in the user’s history.
Specifically, the InfoNCE loss~\cite{oord2018representation} is used during training.
For any prediction $E_{i}^\prime$ in the output sequence of the User LLM, the positive sample is $E_{i}$, and the negative samples are randomly sampled from the dataset excluding the current user sequence. The loss function can be formulated as:
\begin{equation}
    \mathcal{L}_{gen}=-\sum_{j=1}^{b} \sum_{i=2}^{n} \log \dfrac{e^{s(E_{j, i}^\prime, E_{j, i})}}{e^{s(E_{j, i}^\prime, E_{j, i)}} + \sum_{k}^{N} e^{s(E_{j, i}^\prime, E_{j,i,k})}}
\end{equation}
where $s$ is the similarity function with a learnable temperature parameter, $E_{j, i}$ denotes the i-th item embedding produced by the Item LLM in the j-th user's history interaction and $E_{j, i}^\prime$ denotes the i-th item embedding predicted by the User LLM for the j-th user. $N$ is the number of negative samples, $E_{j,i,k}$ represents the k-th negative embedding of $E_{j, i}^\prime$. $b$ represents the total number of users within the batch, $n$ is the length of user history interactions.

\subsubsection{Discriminative Recommendation}

\begin{figure}
    \centering
    \includegraphics[width=\linewidth]{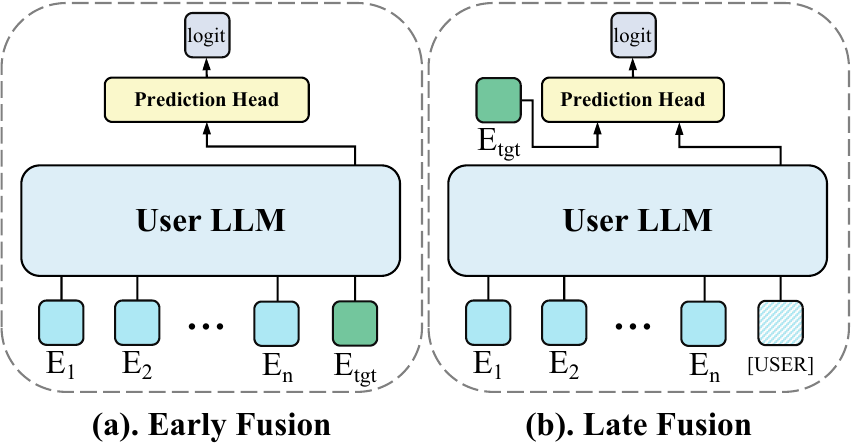}
    \caption{Two User LLM variants for discriminative recommendations.}
    \label{dis}
    \vspace{-0.1in}
\end{figure}

Since discriminative recommendation models still dominate in the industry, we also present an application scheme for HLLM under discriminative recommendation models. The optimization objective of discriminative models is to judge, given a user sequence $U$ and a target item $I_{\text{tgt}}$, whether the user is interested in the target item (e.g., by clicking, liking, purchasing, etc.).

As shown in Figure~\ref{dis}, there are two User LLM variants for discriminative recommendation, while keeping the Item LLM unchanged. \textbf{Early fusion} appends the target item embedding $E_{\text{tgt}}$ to the end of the user's historical sequence, then produces a high-order cross feature through User LLM, and finally inputs this cross feature into the prediction head to generate the final logits.
\textbf{Late fusion}, on the other hand, first uses the User LLM to extract user features, which are independent of the target item, in a manner similar to the Item LLM feature extraction. A special token $\texttt{[USER]}$ is added to the end of the user sequence to extract user representation.
The user embedding and the target item embedding are then input together into the prediction head to predict the final logits.
Early fusion, due to its deep integration of user interests and the target item, tends to perform better but is challenging to apply simultaneously across numerous candidates; conversely, late fusion is more efficient since different candidates share the same user features, but typically sees a performance decline.

The training objective of discriminative recommendation is usually a classification task, such as predicting whether a user will click, etc. For the binary classification example, the training loss is as follows:
\begin{equation}
    \mathcal{L}_{cls} = - \left(y \cdot \log(x) + (1-y) \cdot \log(1-x) \right)
\end{equation}
where $y$ denotes the label of the training sample and $x$ denotes the predicted logit.

Empirically, next item prediction can also be used as an auxiliary loss in discriminative models to further enhance performance. Hence, the final loss can be formulated as follows:

\begin{equation}
    \mathcal{L}_{dis} = \lambda \mathcal{L}_{gen} + \mathcal{L}_{cls}
\end{equation}
where $\lambda$ controls the weight of the auxiliary loss.

\section{Experiments}
In this section, we first introduce the basic experimental settings, and then numerous experiments are conducted to address the following research questions: 

RQ1: Does the general pre-training of the LLM and the fine-tuning with recommendation objectives improve the final recommendation performance? 

RQ2: Does HLLM have good scalability?

RQ3: Are the advantages of HLLM significant compared with other state-of-the-art models?

RQ4: How does the training and serving efficiency compare with ID-based models?

Finally, we demonstrate how to deploy HLLM in online scenarios and achieve real-world benefits.

\subsection{Datasets and Evaluation Setup}

\begin{table}
\centering
\fontsize{9.5}{9.5}\selectfont
\begin{tabular}{@{}lrrr@{}}
\toprule
Dataset   & \#User    & \#Item  & \#Interaction \\ \midrule
Pixel200K & 200,000   & 96,282  & 3,965,656     \\
Pixel1M   & 1,001,822 & 100,541 & 19,886,579    \\
Pixel8M   & 8,886,078 & 407,082 & 158,488,652   \\
Books     & 694,898   & 686,624 & 10,053,086    \\ \bottomrule
\end{tabular}
\caption{Statics of PixelRec and Amazon Book Reviews.}
\label{dataset}
\end{table}

For offline experiments, we evaluate HLLM on two large-scale datasets: PixelRec (including three subsets: 200K, 1M, and 8M)~\cite{pixelrec}, and Amazon Book Reviews (Books)~\cite{books}. Consistent with previous works~\cite{pixelrec, meta}, we adopt the same data preprocessing and evaluation protocols to ensure a fair comparison. A more detailed analysis of these datasets after preprocessing is presented in Table~\ref{dataset} and Figure~\ref{distribution}. We utilize a leave-one-out approach to split the data into training, validation, and testing sets. Performance is measured using the metrics Recall@K (R@K) and NDCG@K (N@K). All open-source datasets are employed solely for training and evaluating in offline experiments.

\begin{table}
\centering
\fontsize{9}{9}\selectfont
\begin{tabular}{@{}llllll@{}}
\toprule
Item LLM   & User LLM   & R@5   & R@10   & N@5  & N@10  \\ \midrule
Scratch    & Scratch    & 3.330 & 5.063 & 2.199 & 2.755 \\ 
Scratch    & \textbf{Pre-trained} & 3.556 & 5.416 & 2.371 & 2.969 \\
\textbf{Pre-trained} & Scratch    & 3.521 & 5.331 & 2.358 & 2.940 \\
\textbf{Pre-trained} & \textbf{Pre-trained} & \textbf{3.755} & \textbf{5.581} & \textbf{2.513} & \textbf{3.100} \\
\bottomrule 
\end{tabular}
\caption{Ablation studies of pre-training on Pixel200K with HLLM-1B.}
\label{RQ1_pre-train}
\vspace{0.1in}
\end{table}

\subsection{Baselines and Training}
For baselines, we use two ID-based sequential recommenders SASRec~\cite{sasrec}, and HSTU~\cite{meta}. They are all aimed at industrial applications and boast state-of-the-art performance.

For offline experiments, the generative recommendation is used to stay consistent with other methods. For the online A/B test, discriminative recommendation is used to better align with the online system\footnote{Experiments demonstrated that most conclusions drawn from the academic dataset still hold true on large-scale industrial benchmarks. }.

In HLLM-1B, we use TinyLlama-1.1B~\cite{tinyllama} for both Item LLM and User LLM. Correspondingly, in HLLM-7B, we utilize Baichuan2-7B~\cite{baichuan2} for both.
Due to resource constraints, HLLMs are trained only 5 epochs on PixelRec and Amazon Reviews while other models are trained 50 and 200 epochs, respectively. The learning rate is set to 1e-4. Each item's text length is truncated to a maximum of 256. 
On PixelRec, following PixelNet~\cite{pixelrec}, we utilize a batch size of 512. The maximum sequence length is set to 10, and the ratio of positive to negative samples is 1:5632. 
On Books, we utilize a batch size of 128, set a maximum sequence length of 50, and the number of negative samples is 512.

For a fair comparison, we also implemented SASRec-1B (replacing its network structure with TinyLlama-1.1B) and HSTU-1B, which uses the same hidden size and number of layers as TinyLlama-1.1B but has only 462M parameters due to the elimination of the traditional FFN.

\subsection{Pre-training and Fine-tuning (RQ1)}

\begin{table}
\centering
\fontsize{9.5}{9.5}\selectfont
\begin{tabular}{@{}llllll@{}}
\toprule
\#Tokens        & R@5   & R@10  & N@5   & N@10  & CSR $\uparrow$         \\ \midrule
0T            & 3.330 & 5.047 & 2.199 & 2.755 & -            \\
0.1T          & 3.539 & 5.142 & 2.399 & 2.915 & 46.11        \\
1T            & 3.613 & 5.409 & 2.414 & 2.993 & 50.22        \\
1T+chat       & 3.610 & 5.387 & 2.411 & 2.984 & 51.36        \\
2T            & 3.650 & 5.510 & 2.466 & 3.063 & 51.64        \\
3T            & \textbf{3.755} & \textbf{5.581} & \textbf{2.513} & \textbf{3.100} & \textbf{52.99}        \\ \bottomrule
\end{tabular}
\caption{The impact of different pre-training token counts on Pixel200K with HLLM-1B. ``+chat'' means SFT on conversation data.
The CSR metric is the average performance on the common sense reasoning tasks.}
\label{pre-training}
\end{table}

\begin{table}
\centering
\fontsize{9.5}{9.5}\selectfont
\begin{tabular}{@{}llllll@{}}
\toprule
Item LLM   & User LLM   & R@5   & R@10   & N@5  & N@10  \\  \midrule
Frozen & \textbf{Learnable} & 0.588 & 0.945 & 0.372 & 0.486 \\
\textbf{Learnable}  & Frozen & 1.619 & 2.470 & 1.070 & 1.343 \\ 
\textbf{Learnable} & \textbf{Learnable} & \textbf{3.755} & \textbf{5.581} & \textbf{2.513} & \textbf{3.100} \\
\midrule
\multicolumn{2}{c}{SASRec-1B}     & 1.973 & 2.868  & 1.352 & 1.640 \\
\bottomrule
\end{tabular}
\caption{Ablation studies of fine-tuning on Pixel200K with HLLM-1B.}
\label{RQ1_fine-tune}
\vspace{-0.1in}
\end{table}

As clearly seen from Table~\ref{RQ1_pre-train}, pre-trained weights are beneficial for HLLM, including both item feature extraction and user interest modeling. Furthermore, as shown in Table~\ref{pre-training}, the performance is positively correlated with the number of pre-trained tokens, indicating that the quality of pre-trained weights also impacts the recommendation task.
However, supervised fine-tuning (SFT) on conversation data can result in slight negative effects, probably because world knowledge is primarily acquired during the pre-training stage, and SFT mainly enhances instruction-following abilities, which do not aid in recommendation tasks~\cite{zhou2024lima}.

It is also evident that fine-tuning both the Item LLM and User LLM is crucial for outperforming ID-based models, as shown in Table~\ref{RQ1_fine-tune}. 
When we freeze the Item LLM and only fine-tune the User LLM, using mean pooling of all token outputs in the last layer of TinyLlama-1.1B as item features, we find that the performance is very poor. This indicates that LLMs trained on predicting the next token are not directly suitable as feature extractors.
Similarly, when we use an Item LLM that has been fine-tuned on Pixel200K and freeze the pre-trained User LLM, the performance remains critically low.

\subsection{Scaling Up (RQ2)}

\begin{table}
\centering
\fontsize{9.5}{9.5}\selectfont
\begin{tabular}{@{}llllll@{}}
\toprule
Item Model & \#Params & R@5   & R@10   & N@5   & N@10  \\ \midrule
BERT-Base  & 110M       & 2.576 & 4.020 & 1.694 & 2.158 \\
BERT-Large & 340M       & 3.032 & 4.635 & 1.993 & 2.508 \\
TinyLlama   & 1.1B       & \textbf{3.484} & \textbf{5.239} & \textbf{2.319} & \textbf{2.883} \\ \bottomrule
\end{tabular}
\caption{Experiments with different sizes of the item model on Pixel200K. SASRec is used as the user model for all.}
\label{item}
\end{table}

\begin{table}[]
\centering
\fontsize{9.5}{9.5}\selectfont
\begin{tabular}{@{}llllll@{}}
\toprule
User Model & \#Params & R@5   & R@10  & N@5   & N@10  \\ \midrule
SASRec     & 4M         & 3.484 & 5.239 & 2.319 & 2.883 \\
Llama-2L  & 0.1B       & 3.494 & 5.233 & 2.338 & 2.898 \\
TinyLlama    & 1.1B       & \textbf{3.521} & \textbf{5.331} & \textbf{2.358} & \textbf{2.940} \\ \bottomrule
\end{tabular}
\caption{Experiments with different sizes of the user model on Pixel200K. Llama-2L maintains the same architecture as Llama but uses only 2 decoder layers. TinyLlama-1.1B is used as the item model for all. All user models are trained from scratch.}
\label{user}
\vspace{-0.1in}
\end{table}

The experimental results for increasing the model's parameter count are shown in Table~\ref{item} and Table~\ref{user}. It can be observed that the growth in the number of parameters for both Item LLM and User LLM consistently leads to performance improvements.
Finally, we scale up both the Item LLM and User LLM from 1 billion parameters to 7 billion parameters on the Amazon Books. As shown in Table~\ref{big}, this leads to further performance improvements, demonstrating that HLLM has excellent scalability.

\begin{figure}
    \centering
    \begin{minipage}{0.49\linewidth}
        \centering
        \includegraphics[width=\textwidth]{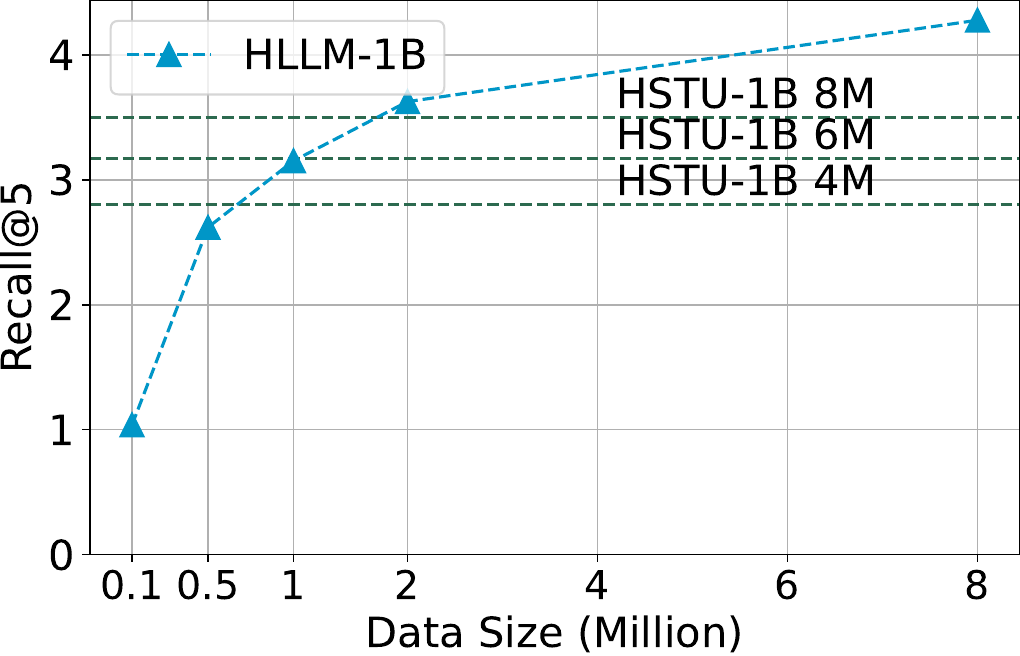}
    \end{minipage}\hfill
    \begin{minipage}{0.49\linewidth}
        \centering
        \includegraphics[width=\textwidth]{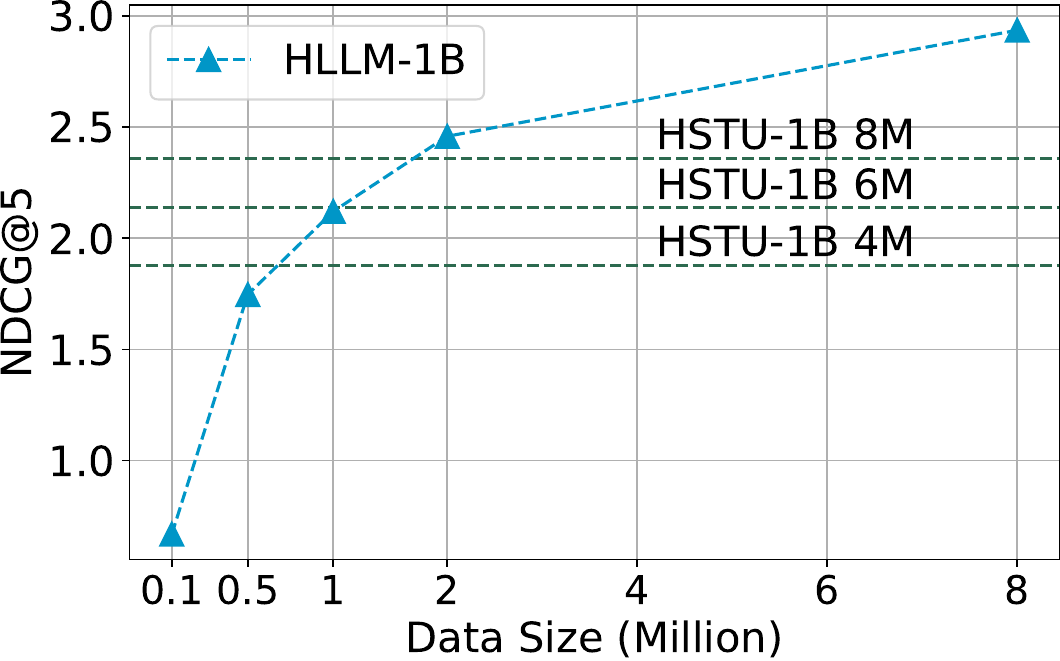}
    \end{minipage}
    \caption{Experiments of HLLM's performance at various data scales. Recall@5 and NDCG@5 are reported.}
    \label{data}
\vspace{-0.1in}
\end{figure}

To explore scalability of data volume, we sampled multiple different scales of data from Pixel8M for training, ranging from 0.1M to 8M in size.
From Figure~\ref{data}, it is evident that HLLM demonstrates remarkable scalability across various data volumes. With increasing data, significant enhancements in performance are observed, and no performance bottlenecks are observed at the current data scale.

We also conducted more comprehensive ablation experiments related to scaling up on a large-scale industrial recommendation dataset to demonstrate the scalability of the HLLM architecture, with detailed experimental results presented in the appendix.

\subsection{HLLM vs. SOTA Methods (RQ3)}

\begin{table*}
\centering
\fontsize{9.5}{9.5}\selectfont
\begin{tabular}{llrrrrrrr}
\toprule
\multicolumn{1}{l}{Dataset}    & \multicolumn{1}{l}{Method} & R@10          & R@50   & R@200  & N@10          & N@50  & N@200 & Impv. (avg) \\ \midrule
\multirow{6}{*}{Pixel8M}    & $\text{SASRec}_\text{vit}$~\shortcite{pixelrec} & 3.589 & - & - &  1.941 & - & - & -27.72\% \\
                            & HSTU$^*$~\shortcite{meta}      & 4.848         & 10.315 & 18.327 & 2.752         & 3.939 & 5.135 & +0.0\% \\
                               & SASRec$^*$  & 5.083  & 10.667 & 18.754 & 2.911  & 4.123 & 5.331 & +3.82\% \\
                               & HSTU-1B$^*$                    & 5.120  & 11.010 & 19.393 & 2.879  & 4.159 & 5.411 & +5.37\% \\
                               & SASRec-1B$^*$                  & 5.142  & 10.899 & 19.044 & 2.915  & 4.166 & 5.383 & +4.83\% \\
                               & HLLM-1B (Ours)  & \textbf{6.129}  & \textbf{12.475} & \textbf{21.179} & \textbf{3.539}  & \textbf{4.919} & \textbf{6.221} & \textbf{+22.93\%}  \\ \midrule
\multirow{12}{*}{Amazon Books} & SASRec~\shortcite{sasrec}   & 3.06  & 7.54   & 14.31  & 1.64  & 2.60  & 3.62 & +0.0\% \\
                               & LEARN~\shortcite{jia2024knowledge}& 4.07 & 9.79   & 18.74  & 2.24 & 3.71  & 4.83 & +34.42\% \\
                               & HSTU-large~\shortcite{meta}     & 4.78  & 10.82  & 19.08  & 2.62  & 3.93  & 5.17 & +47.80\% \\
                               & SASRec$^*$    & 5.35 & 11.91 & 21.02 & 2.98 & 4.40 & 5.76  & +64.96\%      \\
                               & SASRec-1B$^*$ & 5.09 & 11.11 & 19.45 & 2.86 & 4.17 & 5.42  & +55.68\%  \\
                               & HSTU-large$^*$ & 5.00 & 11.29 & 20.13 & 2.78 & 4.14 & 5.47  & +55.61\% \\
                               & HSTU-1B$^*$   & 5.25  & 12.03 & 21.60 & 2.89 & 4.36 & 5.80 & +64.37\%  \\
                               & HLLM-1B (Ours) & \textbf{6.97} & \textbf{14.61} & \textbf{24.78} & \textbf{3.98} & \textbf{5.64} & \textbf{7.16} & \textbf{+108.68\%} \\ \cmidrule(l){2-9}
                               & HSTU-large$^{\dagger*}$ & 6.49  & 12.22  & 19.81     & 3.99   & 5.24          & 6.38    & +88.94\%  \\
                               & HLLM-1B-Scratch$^\dagger$ (Ours)   & 6.85 & 13.95 & 23.19 & 4.02 & 5.56 & 6.95 & +103.65\% \\
                               & HLLM-1B$^\dagger$ (Ours)   & 9.28 & 17.34 & 27.22 & 5.65 & 7.41 & 8.89 & +166.42\% \\
                               & HLLM-7B$^\dagger$ (Ours)   & \textbf{9.39} & \textbf{17.65} & \textbf{27.59} & \textbf{5.69}     & \textbf{7.50}  & \textbf{8.99} & \textbf{+169.58\%} \\ \bottomrule
\end{tabular}
\caption{Performance comparison of HLLM with SOTA models. $\text{SASRec}_\text{vit}$ means SASRec uses the ViT as an image encoder for item encoding and trained by BCE loss from~\cite{pixelrec}. $*$ indicates the result is reproduced by us.
$\dagger$ indicates the number of negative samples and the batch size are increased from 512 and 128 to 28k and 512, respectively.
``Scratch'' indicates both Item LLM and User LLM are trained from scratch.}
\label{big}
\vspace{-0.1in}
\end{table*}

In Table~\ref{big}, we compare the performance of HLLM with the current state-of-the-art models, including ID-based models such as SASRec~\cite{sasrec} and HSTU~\cite{meta}, as well as the text-based model LEARN~\cite{jia2024knowledge} on the Pixel8M and Amazon Book Reviews datasets. They all exhibit excellent performance and are dedicated to industrial practice.

It's clear that HLLM holds a significant performance advantage, decisively outperforming other models on all metrics across all datasets.
Under the same experimental settings, compared to the lowest-performing baseline, HLLM-1B shows an average improvement of \textbf{22.93\%} on Pixel8M, and an even more significant average improvement of \textbf{108.68\%} on Books. In contrast, ID-based models only show a maximum improvement of 5.37\% on Pixel8M and 64.96\% on Books. 

Furthermore, it is notable that when ID-based models increase the number of negative samples and batch size, the performance improvements are relatively modest, especially in R@200 where HSTU-large only increases by \textbf{0.76}, while HLLM-1B increases by \textbf{2.44} under the same setting.
By further increasing the model's parameters, HLLM-7B achieves a significant improvement of \textbf{169.58\%} compared to the baseline, which is highly impressive.

The table also shows that even with fully converged ID-based models, the gains from increasing parameters are minimal. 
On Pixel8M, both SASRec-1B and HSTU-1B show relatively modest improvements compared to smaller sizes, while on Books, SASRec-1B even experiences a decline in all metrics.
In contrast, for HLLM, scaling up from HLLM-1B to HLLM-7B still results in corresponding performance improvements on recommendation tasks, demonstrating the superiority of the HLLM architecture.

\subsection{Training and Serving Effeciency (RQ4)}
\begin{table}
\centering
\fontsize{9.5}{9.5}\selectfont
\begin{tabular}{@{}lllll@{}}
\toprule
Method & R@5   & R@10  & N@5   & N@10  \\ \midrule
HSTU-1B    & 3.501 & 5.120 & 2.358 & 2.879 \\
$\text{HLLM-1B}_{\text{cache}}$    & 3.585 & 5.218 & 2.432 & 2.958 \\
HLLM-1B    & \textbf{4.278} & \textbf{6.106} & \textbf{2.935} & \textbf{3.524} \\ \bottomrule
\end{tabular}
\caption{Experiments on the effectiveness of item caching. $\text{HLLM-1B}_{\text{cache}}$ utilizes a pre-trained item HLLM to extract item features, but the parameters are frozen.}
\label{freeze}
\vspace{-0.1in}
\end{table}

Firstly, HLLM shows better training data efficiency than ID-based models. As shown in Figure~\ref{data}, HLLM requires only one-sixth to one-fourth of the data volume to achieve performance on par with ID-based methods.

Previous extensive experiments have shown that fully fine-tuning the entire HLLM significantly improves performance but requires real-time encoding of all items during inference, which is inefficient. Thanks to the decoupling of item and user encoding in HLLM, our architecture can reduce computational complexity by caching item embeddings in advance. 
To demonstrate the feasibility of item caching, we pre-trained HLLM on sequences longer than 10 from the Pixel8M dataset, truncating sequences at the tenth position to avoid data leakage, covering 3 million users. Based on this pre-trained HLLM, we freeze the Item LLM and fine-tune only the User LLM on Pixel8M. Results in Table~\ref{freeze} show that while freezing the Item LLM leads to some metric decreases, performance still exceeds ID-based models, proving item caching is more effective. Given that user behaviors in industrial scenarios far exceed the number of items, HLLM's training and serving costs can match those of ID-based models.
Notably, our pre-training data constitutes less than half of Pixel8M, with some items not appearing in the pre-training data, yet we still achieve respectable performance. Experiments on industrial data show that as the amount of pre-training data increases, the gap between the item caching and the full fine-tuning is largely narrowed.

\subsection{Online A/B Test}

Apart from offline experiments, HLLM is also targeted at and successfully applied in real-world industrial practices. 
For simplicity, flexibility, and to better align with the online system, we adopted HLLM-1B, using the discriminative recommendation approach with the late fusion variant for optimization.
Considering the balance between performance and efficiency, our training process is divided into the following three stages:

Stage \Rmnum{1}: End-to-end training of all HLLM parameters, including Item LLM and User LLM with discriminative loss. 
The user history sequence length is truncated to 150 to accelerate training.

Stage \Rmnum{2}: We first use the Item LLM trained in Stage \Rmnum{1} to encode and store the embeddings of all items in the recommendation system. 
We then continue to train only the User LLM by retrieving the necessary item embeddings from storage. 
Since this stage only trains the User LLM, it significantly reduces the training demand, allowing us to extend the user sequence length from 150 to 1000, further enhancing the effectiveness of the User LLM.

Stage \Rmnum{3}: After extensive data training in the first two stages, the HLLM model parameters are no longer updated. We extract features for all users which are then combined with item LLM embeddings and other existing features and fed into the online recommendation model for training.

\begin{figure}
    \centering
    \includegraphics[width=\linewidth]{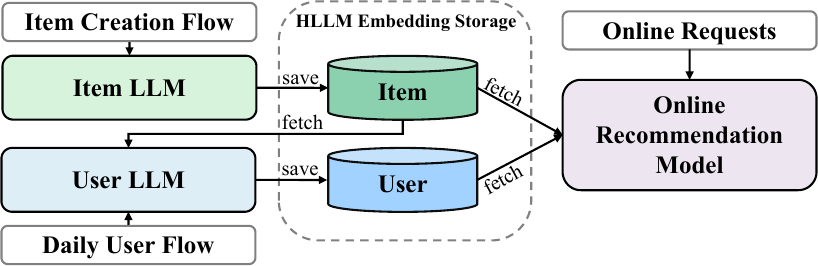}
    \caption{An overview of the online system.}
    \label{online}
\vspace{-0.2in}
\end{figure}

Regarding serving, as shown in Figure~\ref{online}, item embeddings are extracted when they are created, and user embeddings are updated on a daily basis only for users who had activity the previous day. Embeddings of items and users are stored for online model training and serving. Under this approach, the inference time of the online recommendation system is virtually unchanged.

Finally, we test HLLM in online A/B experiments of the ranking task. Key metrics have shown a significant increase of \textbf{0.705\%}.

\section{Conclusion}
In this paper, we propose a novel Hierarchical Large Language Model (HLLM) architecture designed to enhance sequential recommendations. HLLM leverages LLMs to extract item features and model user interests, effectively integrating pre-training knowledge into the recommendation system, and it is proved that fine-tuning with recommendation objectives is essential. HLLM exhibited excellent scalability with larger model parameters. Experiments demonstrated that HLLM outperforms traditional ID-based models, achieving state-of-the-art results on academic datasets. Real-world online A/B testing further validated HLLM's practical efficiency and applicability, marking a significant advancement in the field of recommendation systems.

\bibliography{aaai25}

\clearpage
\appendix
\section{More Experiments on Academic Datasets}
\subsection{Textual Input Length and Richness of Item LLM}
\begin{table}
\centering
\fontsize{9.5}{9.5}\selectfont

\begin{tabular}{ccclll}
\toprule
Tag & Title &Description & Length & R@5   & R@10  \\ \midrule
\usym{2713} &             &             & 64     & 0.082 & 0.196 \\ 
            & \usym{2713} &             & 64     & 3.520 & 5.317 \\ 
\usym{2713} & \usym{2713} &             & 64     & 3.610 & 5.439 \\
\usym{2713} & \usym{2713} & \usym{2713} & 64     & 3.647 & 5.502 \\
\usym{2713} & \usym{2713} & \usym{2713} & 256    & \textbf{3.755} & \textbf{5.581} \\ \bottomrule
\end{tabular}
\begin{tabular}{ccclll}
\toprule
Tag & Title & Description & Length & N@5   & N@10  \\ \midrule
\usym{2713} &             &             & 64     & 0.049 & 0.086 \\ 
            & \usym{2713} &             & 64     & 2.348 & 2.926 \\ 
\usym{2713} & \usym{2713} &             & 64     & 2.415 & 3.003 \\
\usym{2713} & \usym{2713} & \usym{2713} & 64     & 2.430 & 3.026 \\
\usym{2713} & \usym{2713} & \usym{2713} & 256    & \textbf{2.513} & \textbf{3.099} \\ \bottomrule
\end{tabular}
\caption{Ablation studies of text length and richness on Pixel200K.}
\label{text-length}
\end{table}

By default, we input all types of text information with a length of 256. Here, we conduct ablation experiments on text length and richness.
Table~\ref{text-length} shows that the text content has a significant impact on the final performance. Richer text content and longer text lengths allow the Item LLM to extract more detailed item features, better differentiate between items, and more effectively aid the User LLM in modeling user interests.

\subsection{Method of Item LLM Feature Extraction}

\begin{table}
\centering
\fontsize{9.5}{9.5}\selectfont
\begin{tabular}{@{}lllll@{}}
\toprule
Method & R@5   & R@10  & N@5   & N@10  \\ \midrule
Mean Pooling    & 3.386 & 5.159 & 2.257 & 2.826 \\
$\texttt{[ITEM]}$ Token & \textbf{3.484} & \textbf{5.239} & \textbf{2.319} & \textbf{2.883} \\ \bottomrule
\end{tabular}
\caption{Ablation studies of Item LLM feature extraction method on Pixel200K. Mean pooling refers to using the mean pooling of hidden states from the last layer of the Item LLM as the item features. SASRec is used as the user model.}
\label{meanpooling}
\end{table}

To enable LLMs trained on next token prediction to have feature extraction capabilities, we add a special token $\texttt{[ITEM]}$ at the end of the text input. Another feasible feature extraction approach is to take the average of the hidden states from the final layer of the LLM to represent the features of the entire sentence. Table~\ref{meanpooling} shows the comparison results of these two methods.
As can be seen, using the $\texttt{[ITEM]}$ token is better than mean pooling.

\subsection{Sequence Length of User LLM}

We explore the impact of input sequence length of User LLM on HLLM's recommendation performance in Table~\ref{seqlength}. Similar to other sequential recommenders, HLLM can also benefit from expanding the length of the input sequence. Although the table shows only modest performance gains with increasing sequence length, we suspect this is likely because user sequence lengths are generally quite short in the academic dataset as shown in Figure~\ref{distribution}. 
As shown in Appendix~\ref{industry}, in the real-world industrial systems, where user behavior sequences are typically very long, extending the sequence length allows HLLM to achieve stable performance improvement.

\begin{table}[t]
\centering
\fontsize{9.5}{9.5}\selectfont

\begin{tabular}{cllll}
\toprule
Length & R@5  & R@10 & R@50 & R@200  \\ \midrule
10    & 5.201 & 7.564 & 16.220 & 28.776 \\
30    & 5.235 & 7.605 & 16.293 & 28.837 \\
50    & \textbf{5.238} & \textbf{7.631} & \textbf{16.416} & \textbf{28.959} \\ \bottomrule
\end{tabular}

\begin{tabular}{cllll}
\toprule
Length & N@5  & N@10 & N@50 & N@200  \\ \midrule
10    & 3.538 & 4.299 & 6.176 & 8.052 \\
30    & 3.556 & 4.319 & 6.205 & 8.081 \\
50    & \textbf{3.568} & \textbf{4.338} & \textbf{6.244} & \textbf{8.119} \\ \bottomrule
\end{tabular}
\caption{Experiments on the sequence length of User LLM on Pixel1M.}
\label{seqlength}
\end{table}

\begin{table}[t]
\centering
\fontsize{9.5}{9.5}\selectfont
\begin{tabular}{lllll}
\toprule
Input Features     & R@5   & R@10  & N@5   & N@10  \\ \midrule
Item ID         & 4.105 & 6.082 & 2.773 & 3.409 \\
LLM Emb          & 5.201 & 7.564 & 3.538 & 4.299 \\
LLM Emb + Item ID   & 5.154 & 7.501 & 3.504 & 4.260 \\
LLM Emb + Timestamp & \textbf{5.779} & \textbf{8.319} & \textbf{3.953} & \textbf{4.770} \\ \bottomrule
\end{tabular}
\caption{Ablation studies of input features of User LLM on Pixel1M. LLM Emb represents the item features extracted using the Item LLM based on textual descriptions.}
\label{addfeature}
\end{table}

\subsection{Compatibility with ID-based Features}

In the previous sections, we primarily modeled item and user features based on the textual descriptions of items. Most current recommendation systems, however, still rely on ID features, including not only Item IDs but also features like actions, timestamps, and item categories in ID form. Here, we present a compatibility solution for integrating HLLM with ID features, and demonstrate that complementary ID features, when combined with item descriptions, can indeed bring significant improvements to HLLM, further highlighting its application value in industrial environments.

Here, we choose the raw item IDs and timestamps as ID features for validation.
The item IDs are transformed into id embeddings through an embedding lookup table.
The behavior's timestamp is first split into specific year, month, day, hour, minute, and second components, obtaining the timestamp embedding as Algorithm~\ref{alg:code}.
We perform sum pooling with the ID features and item LLM embeddings before inputting them into the User LLM.
The prediction target during training remains the item embedding extracted by the Item LLM, and the experimental results are shown in Table~\ref{addfeature}.
The introduction of item IDs actually results in a slight decrease in performance, likely because the item IDs do not provide incremental information beyond what is already captured by the textual descriptions, which comprehensively describe the item's characteristics and are sufficiently extracted by the Item LLM.
However, the improvement resulting from the introduction of timestamps is very pronounced, as timestamps complement the textual descriptions. This also demonstrates that our method can be compatible with ID-based features.

\section{Scaling Up of HLLM on Industrial Dataset}
\label{industry}
More extensive experiments are conducted on a large-scale industrial dataset to evaluate the scalability of HLLM. 

Douyin has a vast number of users and recommendation candidates, with extensive records of user behavior. We construct a dataset comprising 30 million samples from the past 3 years' logs. Each sample includes only the user's historical click sequence, the target item, and a label indicating whether the item was clicked or not. We validate the effectiveness of HLLM in a discriminative recommendation system, using AUC as the evaluation metric, and verifying scalability from two aspects: the sequence length of User LLM, and the parameters of both Item LLM and User LLM.

\begin{table}[t]
\centering
\begin{tabular}{ll}
\toprule
Sequence Length & AUC  \\ \midrule
200             & 0.7429  \\
500             & 0.7446  \\
1,000           & \textbf{0.7458} \\ \bottomrule
\end{tabular}
\caption{Experiments on the sequence length of User LLM on the industrial dataset.}
\label{seqlength-douyin}
\end{table}

\begin{table}[t]
\centering
\begin{tabular}{lll}
\toprule
Item LLM & User LLM & AUC  \\ \midrule
1B       & 1B       & 0.7458  \\
1B       & 7B       & 0.7498  \\
7B       & 1B       & 0.7517  \\
7B       & 7B       & \textbf{0.7533} \\ \bottomrule
\end{tabular}
\caption{Experiments with different sizes of Item LLM and User LLM on the industrial dataset.}
\label{scale}
\end{table}

\subsection{Sequence Length of User LLM}
\begin{figure}
    \centering
    \begin{minipage}{ \linewidth}
        \centering
        \includegraphics[width=\textwidth]{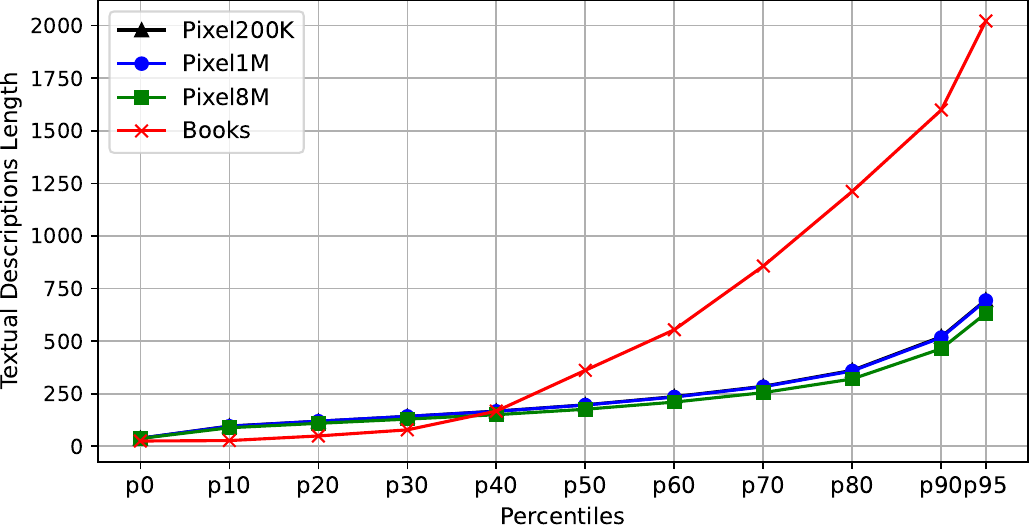}
    \end{minipage}\hfill
    \begin{minipage}{\linewidth}
        \centering
        \includegraphics[width=\textwidth]{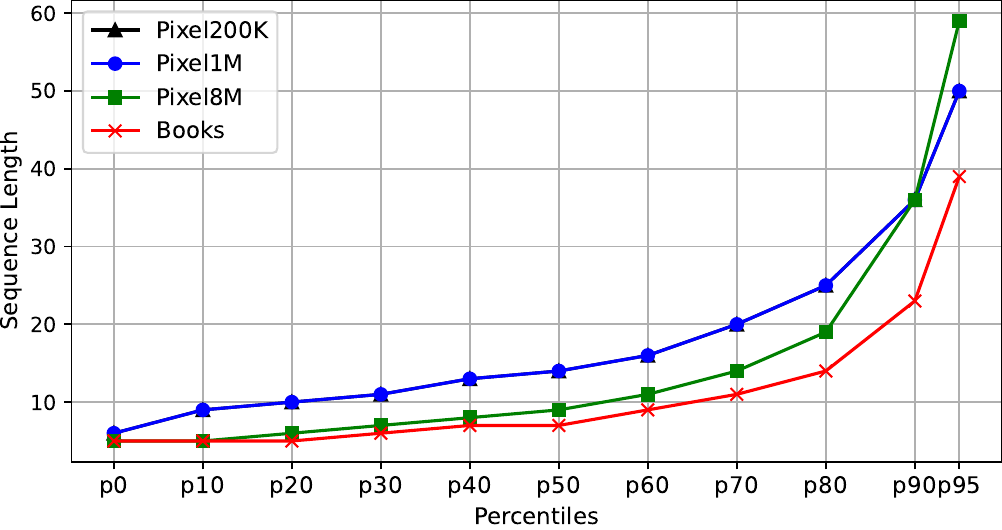}
    \end{minipage}
    \begin{minipage}{\linewidth}
        \centering
        \includegraphics[width=\textwidth]{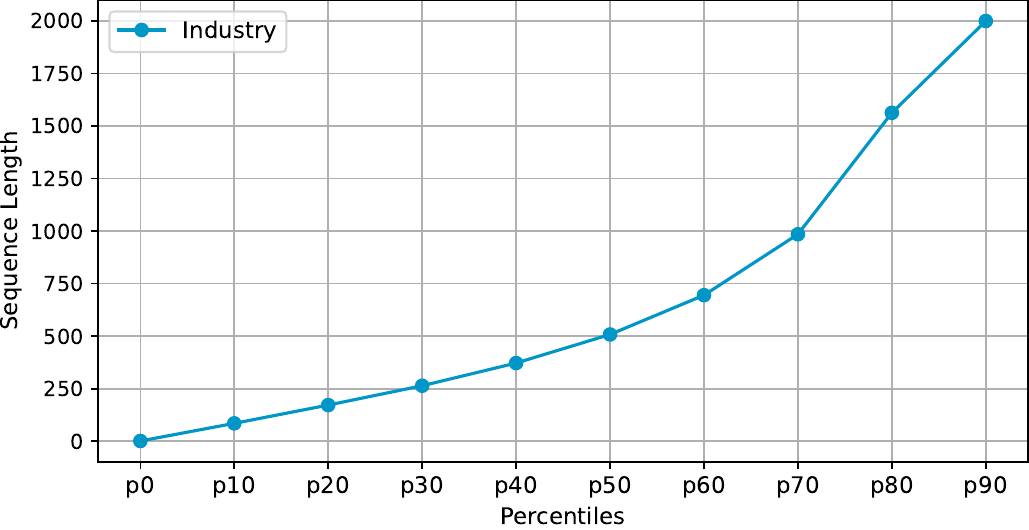}
    \end{minipage}
    \caption{Distribution of textual descriptions (flattening all attributes) and sequence lengths in Pixel200K, Pixel1M, Pixel8M, Amazon Book Reviews and industrial scenario. Since Pixel200K is randomly sampled from Pixel1M, their distributions are consistent. We truncate the sequence length to a maximum of 2,000 for industrial data, hence p90 is exactly 2,000.}
    \label{distribution}
\end{figure}

The length of user behavior sequences in the industrial dataset is shown in Figure~\ref{distribution}.
And table~\ref{seqlength-douyin} shows the impact of user sequence length, with HLLM's performance steadily increasing as the sequence length grows. This illustrates HLLM's substantial potential in modeling users with longer sequences.

\subsection{Parameters of Item LLM and User LLM}
Table~\ref{scale} illustrates the impact of the parameters of HLLM in industrial scenario. For both Item LLM and User LLM, AUC consistently increases with the growth in the number of parameters.

\begin{algorithm*}
\caption{Pseudo code of timestamp processing in a PyTorch-like style.}
\label{alg:code}
\definecolor{codeblue}{rgb}{0.25,0.5,0.5}
\definecolor{codegreen}{rgb}{0,0.6,0}
\definecolor{codegray}{rgb}{0.5,0.5,0.5}
\definecolor{codepurple}{rgb}{0.58,0,0.82}
\definecolor{backcolour}{rgb}{0.95,0.95,0.92}
\definecolor{codered}{rgb}{0.9,0,0}
\lstset{
  backgroundcolor=\color{white},
  basicstyle=\fontsize{7.5pt}{7.5pt}\ttfamily\selectfont,
  columns=fullflexible,
  breaklines=true,
  captionpos=b,
  commentstyle=\fontsize{7.5pt}{7.5pt}\color{codeblue},
}
\begin{lstlisting}[language=python]
class TSEmbedding(nn.Module):
    def __init__(self, time_num=6, time_dim=512, user_dim=2048):
        super().__init__()
        # Control the precision of time, such as 4 to the hour, and 6 to the second.
        self.time_num = time_num
        self.time_embeddings = nn.ModuleList(nn.Embedding(x, time_dim) for x in [2100, 13, 32, 24, 60, 60])
        # Projection from time_dim to user_dim
        self.merge_time = MLP(time_dim * time_num, user_dim)
    
    def split_time(self, timestamps: List) -> List:
        # Split timestamps into specific components.
        # (seq) -> (seq, 6)
        split_time = []
        for time in timestamps:
            dt = datetime.datetime.fromtimestamp(time)
            split_time.append([dt.year, dt.month, dt.day, dt.hour, dt.minute, dt.second])
        return split_time
    
    def forward(self, timestamps: List) -> torch.tensor:
        # Times: timestamp of each item in List format (bs, seq)
        # (bs, seq) -> (bs, seq, 6)
        time_seq = torch.tensor([self.split_time(x) for x in timestamps])
        # (bs, seq, 6) -> [(bs, seq, time_dim)] * time_num
        time_emb = [self.time_embeddings[i](time_seq[...,i]) for i in range(self.time_num)]
        # [(bs, seq, time_dim)] * time_num -> (bs, seq, time_dim * time_num)
        time_emb = torch.cat(time_emb, dim=-1)
        # (bs, seq, time_dim * time_num) -> (bs, seq, user_dim)
        time_emb = self.merge_time(time_emb)
        return time_emb
\end{lstlisting}
\end{algorithm*}

\end{document}